\documentclass[twocolumn,twoside]{IEEEtran}
\usepackage{amsmath,graphicx}

\author{Tommaso Toffoli\thanks{Tommaso Toffoli (\ttt{tt\char"40bu.edu}), ECE
Department, Boston University, 8 Saint Mary's St., Boston, MA 02215.}}

 \title{Thermodynamics of used punched tape:
	A weak and a strong equivalence principle}

\let\ttt\texttt
\let\tsc\textsc

 \makeatletter

 \let\tdf=\textbf

 \DeclareRobustCommand\em
        {\@nomath\em \ifdim \fontdimen\@ne\font >\z@
                       \upshape \else \slshape \fi}

\def\@begintheorem#1#2{\sl \trivlist \item[\hskip \labelsep{\bf #1\ #2}]}
\def\@opargbegintheorem#1#2#3{\sl \trivlist
     \item[\hskip \labelsep{\bf #1\ #2\ (#3)}]}

 \mathchardef\BY="0202
 \def\@empty{}
 \newcommand{\asin}[2][]{{%extra braces for lower level, to forget \@cite
    \def\t@mp{#1}%
    \def\@cite##1##2{\marginpar{\hfil{\footnotesize$
    \ifx\t@mp\@empty\text{##2}\else\frac{\text{##1}}{\text{##2}}\fi$}\hfil}}%
\cite[#1]{#2}}}

 \def\pages#1{}	

 \newcommand{\sectlabel}[1]{\label{sect:#1}}
 \newcommand{\footlabel}[1]{\label{foot:#1}}
 \newcommand{\eqlabel}[1]{\label{eq:#1}}
 \newcommand{\figlabel}[1]{\label{fig:#1}}

 \newcommand{\Chapt}[2][]{\def\t@mp{#1}%
\chapter{#2} \ifx\t@mp\@empty\else\sectlabel{#1}\fi}
 \newcommand{\Sect}[2][]{\def\t@mp{#1}%
\section{#2} \ifx\t@mp\@empty\else\sectlabel{#1}\fi}
 \newcommand{\Subsect}[2][]{\def\t@mp{#1}%
\subsection{#2} \ifx\t@mp\@empty\else\sectlabel{#1}\fi}
 \newcommand{\Foot}[2][]{\def\t@mp{#1}%
\footnote{#2} \ifx\t@mp\@empty\else\footlabel{#1}\fi}
 \newcommand{\Eq}[2][]{\def\t@mp{#1}%
\begin{equation}#2\ifx\t@mp\@empty\notag\else\eqlabel{#1}\fi\end{equation}}
 \newcommand{\Eqaligned}[2][]{\def\t@mp{#1}%
\begin{equation}\begin{aligned}#2\end{aligned}
\ifx\t@mp\@empty\notag\else\eqlabel{#1}\fi
\end{equation}}

 \newcommand{\sect}[1]{\S\ref{sect:#1}}      % Ref. to section or subsection
 \newcommand{\eq}[1]{(\ref{eq:#1})}	% Ref. to equation
 \newcommand{\fig}[1]{Fig.~\ref{fig:#1}}

 \long\def\endsubsection#1{\smallskip\hbox to\hsize{\leaders\hrule\hfill\ \sect{#1}}\medskip}

  \def\@arabic#1{\number #1} % my redefinition

\def\figstrut#1{\hbox to\linewidth{\vrule height#1\hfill}}

\newcommand{\Fig}[3][]{% [label], picture, caption
\begin{figure}[!htb]
 \centering{\leavevmode#2}%
 \caption{#3}
 \figlabel{#1}
\end{figure}                 }

 \newcommand{\Figwide}[3][]{% [label], picture, caption
 \begin{figure*}[!t]
  \centering\leavevmode#2%
  \caption{#3}
  \figlabel{#1}
 \end{figure*}                 }

\def\cstrip#1{\setbox0=\hbox{$#1$}\kern-.5\wd0\lower2pt\box0}
\def\rstrip#1{\setbox0=\hbox{$#1$}\kern-\wd0\lower2pt\box0}
\def\lstrip#1{\setbox0=\hbox{$#1$}\lower2pt\box0}
\def\tstrip#1{\setbox0=\hbox{$#1$}\kern-.5\wd0\lower\ht0\box0}
\def\bstrip#1{\setbox0=\hbox{$#1$}\kern-.5\wd0\raise\ht0\box0}
\def\Lstrip#1{\setbox0=\hbox{$\mskip2mu#1$}\lower2pt\box0}

\def\idpad{\thinspace}
\def\id{\idpad\begingroup \tt \let\do\@makeother \dospecials 
          \@ifstar{\@sid}{\@id}}
\def\@sid#1{\def\@tempa ##1#1{##1\endgroup\idpad}\@tempa}
\def\@id{\obeyspaces \frenchspacing \@sid}

\makeatother

 \def\above#1#2#3{\genfrac{}{}{0pt}{#1}{#2}{#3}}

\def\compress{\itemsep=2pt \topsep=2pt \parsep=0pt \partopsep=0pt}

\def\LN{\mathop{\overline{\smash{\mathrm{ln}}\vphantom x}}\nolimits}
\def\der#1#2{\frac{d#2}{d#1}}

\def\overbar#1{\kern.05em\overline{#1}\kern.05em}

\def\pbar{\overbar{p}}
\def\qbar{\overbar{q}}
\def\qhat{\widehat{q}}
\def\pp{{p'}}
\def\ppbar{\overbar{\pp}}
\def\pphat{\widehat{\pp}}
\def\bit{\mathrm{bit}}
\def\info{I}
\def\mutu{{\Delta\info}}

\begin{document}

\maketitle

 \begin{abstract}We study the repeated use of a \emph{monotonic} recording
medium---such as punched tape or photographic plate---where marks can be
added at any time but never erased. (For practical purposes, also the
electromagnetic ``ether'' falls into this class.) Our emphasis is on the case
where the successive users act independently and selfishly, but not
maliciously; typically, the ``first user'' would be a blind natural process
tending to degrade the recording medium, and the ``second user'' a human
trying to make the most of whatever capacity is left.

\smallskip To what extent is a length of used tape ``equivalent''---for
information transmission purposes---to a shorter length of virgin tape? Can
we characterize a piece of used tape by an appropriate ``effective length''
and forget all other details? We identify two equivalence principles. The
\emph{weak} principle is exact, but only holds for a sequence of
infinitesimal usage increments. The \emph{strong} principle holds for any
amount of incremental usage, but is only approximate; nonetheless, it is
quite accurate even in the worst case and is virtually exact over most of the
range---becoming exact in the limit of heavily used tape.

\smallskip The fact that strong equivalence does not hold \emph{exactly}, but
then it does \emph{almost} exactly, comes as a bit of a surprise.
 \end{abstract}


Bob---a poor computer science student---has found, rummaging through Alice's
dump, a large amount of used punched tape ``in good conditions''. He doesn't
care for the data that is already on the tape: he would like to reuse the
tape for storing his own data. He wants to be able to use a standard tape
read/punch unit, which can sense holes in the tape and punch new ones but not
remove holes that are already there. Since holes already made cannot be
undone, the storage density Bob can expect to achieve is less than with
virgin tape, and will depend on the actual conditions of the tape.

To what extent is a length of used tape ``equivalent''---for information
transmission purposes---to a shorter length of virgin tape?  Are there any
qualitative differences between tapes that have been used to different
degrees, or can one characterize a piece of used tape simply by its
``effective length'' and forget all other details?

\medskip

The theme we develop is complementary to that of Rivest and
Shamir\cite{rivest-worm} (also cf.\ \cite{maier}). They stress the
information-engineering aspects of reusing a tape generated by a cooperative
partner in a pre-planned context. On the other hand, we are interested in a
situation where the other party, while presumed non-malicious, volunteers no
cooperation and pursues independent goals (if any goals can be made out);
what we typically have in mind for ``the other party'' is \emph{natural
processes}.\footnote
 {As humans become more proficient at exploiting physical mechanisms on a finer
and finer scale for computational purposes, computation will look more and
more like an attempt to encroach on a turf already jammed near capacity by
heavy ``native'' traffic---the near-equilibrium bustle of microscopic matter
(cf.~Dyson\cite{dyson}). The present study is part of a wider program aimed at
exploring this kind of computational regime.}

The cumulative channel capacity of randomly-punched used tape was first
investigated in \cite{wolf} (also see references therein), some of whose
results we simplify and extend.  References \cite{dolev} and \cite{heegard}
discuss coding algorithms that dynamically adjust to ``stuck-at-0'' faults on
the tape (cells that will not punch) sensed during punching, and
``stuck-at-1'' faults sensed during or before punching. A paper related to
the present one in spirit if not in detailed substance is ``Writing on dirty
paper'' by Costa\cite{costa}, whose moral (``Do the best with what you
have'') we make our own.

\medskip

If you have no time at all, read just \sect{dialogues}---a self-contained,
intuitive debriefing.

\Sect[orientation]{Orientation}

Each position on the tape where a hole may appear is called a \tdf{cell}; the
two possible cell states are \tdf{hole} and \tdf{blank}. The instructions to
the punch unit are \tdf{punch} and \tdf{spare}, with the following results on
the tape
 $$
	\begin{array}{llll}
		\hbox{\small\sc old state} &\hbox{\small\sc action}& &\hbox{\small\sc new state}\\\hline
		\hbox{blank} & \hbox{spare} & \mapsto & \hbox{blank}\\
		\hbox{hole} & \hbox{spare}  & \mapsto & \hbox{hole}\\
		\hbox{blank} & \hbox{punch} & \mapsto & \hbox{hole}\\
		\hbox{hole} & \hbox{punch}  & \mapsto & \hbox{hole}
	\end{array}
 $$

A hole (or punch) distribution that factors into identical independent
distributions for the individual cells---and is thus characterized by a
single number, namely, the hole (or punch) density---will be called
\tdf{canonical}. We shall assume that on each round of usage or \tdf{stage}
the tape starts with a canonical hole distribution of density $p$ and comes
out with a uniform hole density $p'$; furthermore, we assume that the
intervening punching process packs on the tape the \emph{maximum} amount of
new information compatible with startng density $p$ and target
density $p'$. According to Shannon's theorem, such maximum efficiency can
asymptotically be achieved by means of sufficiently long block codes. From
the above assumptions one can prove that both the punch distribution $q$
yielded by an optimal code and the resulting hole distribution $p'$ must be
canonical as well. Thus, our usage assumptions imply that, starting from
virgin tape---whose distribution is, of course, canonical with $p=0$---input,
punch, and output distributions will be canonical at every successive
stage. For this reason, in what follows all distributions will be tacitly
understood to be canonical.

 The result of applying a punch density $q$ to a hole
density $p$ is a new hole density
 \begin{equation}\eqlabel{pprime}
	\pp=1-(1-p)(1-q).
 \end{equation}
	A canonical punch distribution entails that, once the input hole
density $p$ is known, there is no further advantage in knowing the position
of the individual holes; in other words, overpunching can be carried out in a
\emph{data-blind} fashion.

\medskip

Let's examine a few distinguished cases.
 \begin{list}{$\bullet$}{\compress}
	\item If $p=0$ the tape is blank---Bob can resell it as virgin
tape.
	\item If $p=1-p=1/2$, the tape has already been utilized by Alice at
its maximum information capacity of one bit per cell. That would seem to
leave Bob with no room for further information storage. But remember that he
doesn't care about the old information: punching new holes will destroy some
of it but will encode some of his own! In fact (see \sect{monotonic channel}
below), with a punch density $q=3/5$, Bob can record on the tape as much as
about .322 bits per cell.
	\item If $p=1$, the tape carries no information for Alice---just as
in the case $p=0$. However, now there is no way Bob can put any information
on it.  Alice wantonly spoiled the tape.
 \end{list}

\Sect[notation]{Notation}

If $p$ is a probability, it will be convenient to write $\pbar$ for
$1-p$. Thus, in \eq{pprime},
 $$
	\pp=1-\pbar\qbar=\overline{\pbar\qbar},\quad \text{or}\quad 
	\qbar=\ppbar/\pbar.
 $$

We shall use natural logarithms throughout. It will be convenient to write
$\LN x$ for $-\ln x$. The \tdf{self-information function}, defined as
 $$
	y = x\LN x,
 $$
	will play an important role in the equivalence principles discussed
here (see \sect{strong-principle}). The \tdf{binary entropy function},
defined by
 $$
	H(p)=p\LN p+\pbar\LN\pbar,
 $$
	is the average of the self-information function over the binary
distribution $\{p,\pbar\}$.

Both self-information and binary entropy, as defined here, measure
information in natural units or \tdf{nats}. Conversion of information
quantities to binary units or \tdf{bits} is achieved by explicitly factoring
out the constant
 $$
	\bit=\ln 2\approx.693;
 $$ 
	thus, for example, the entropy of four equally probable messages
is $\ln4=2\ln2 =2\ \bit$.

If $X$ and $Y$ are random variables, $P(x)$ will denote the probability that
$X=x$, and $P(x.y)$ the probability that $X=x$ and $Y=y$.
The \tdf{mutual information} between $X$ and $Y$ is defined as
 $$
	\{X;Y\}	= \sum_{x,y}P(xy)\LN\frac{P(x)P(y)}{P(x.y)}.
 $$

For more background on information theory, see the excellent introduction by
Abramson\cite{abramson}.

\Sect[monotonic channel]{Used tape as a monotonic binary channel}

Under the above assumptions (\sect{orientation}), used punched tape may be
viewed as a communication channel affected by monotonic noise.  In the
channel diagram of \fig{mono_channel-diagram}, the input variable $X$
represents the instruction given to the punch unit while scanning a cell, and
the output variable $Y$ represents the resulting cell state.  An ``error''
occurs when a cell spared by the punch unit turns out already to contain a
hole. The conditional probability $P(\hbox{hole}|\hbox{spare})$ associated
with this transition equals the current hole density $p$.

\Fig[mono_channel-diagram]{\input mono_channel.pic
}{Channel diagram
of used punched tape viewed as a monotonic-error binary channel.}

From the joint and marginal distributions of $X$ and $Y$, namely,
 \Eq[mono_channel-matrix]{
 \begin{array}{c|c|c|c|}
 \multicolumn{1}{c}{}&\multicolumn{1}{c}{y}&
				\multicolumn{1}{c}{\hbox{blank}}&\multicolumn{1}{c}{\hbox{hole}}\\
 \cline{3-4}
 \multicolumn{1}{c}{x}&\multicolumn{1}{c|}{}&
	\pbar\,\qbar & 1-\pbar\,\qbar\\\cline{2-4}
 \hbox{spare} & \qbar & \pbar\,\qbar   & p\qbar\\\cline{2-4}
 \hbox{punch} & q & 0              & p\\\cline{2-4}
 \end{array},
}
	one obtains, for this channel operated
at a punch density $q$, a mutual information
 \Eq[mono_mutu]{
       \mutu = H(\pbar\,\qbar)-\qbar H(p)
		= H(\pp)-\frac\ppbar\pbar H(p).
}
	The quantity $\mutu$ is the amount of \emph{new} information that
can be encoded on a tape having a hole density $p$ by punching it with a
density $q$, resulting in a new hole density $\pp(p,q)=1-\pbar\qbar$.

 \emph{The relation expressed by equation \eq{mono_mutu}---plotted in
\fig{mono_mutu}---completely characterizes
the bulk properties of punched tape as a communication channel.  The rest of
this paper is devoted to extracting some of its implications.}

\Fig[mono_mutu]{%
 \begin{picture}(248,156)(-24,-14)
 \small
  \put(0,0){\includegraphics{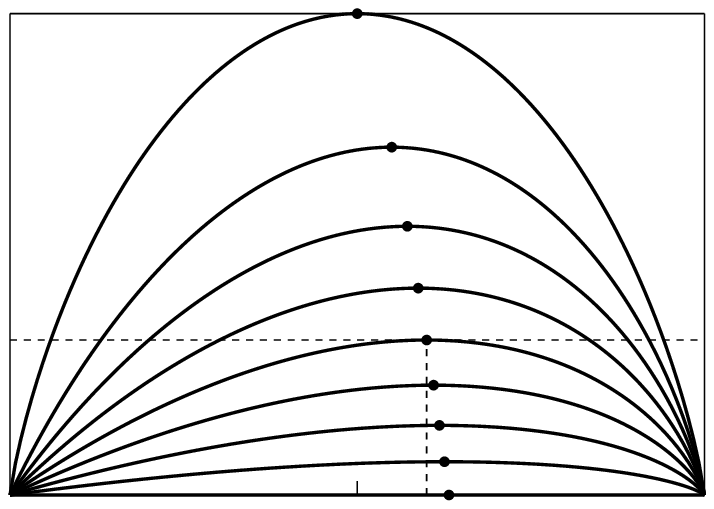}}
 \put(224,-10){\rstrip{\hbox{\normalsize$q\to$}}}
 \put(0,-8){\cstrip{0}}
 \put(100,-8){\cstrip{\frac12}}
 \put(118,-8){\cstrip{\frac35}}
 \put(132,-8){\cstrip{1{-}\frac1e}}
 \put(200,-8){\cstrip{1}}
 
 \put(-24,92){\lstrip{\hbox{\normalsize$\above0\uparrow\mutu$}}}
 \put(201,129){\lstrip{(\negthinspace\times\hbox{bit}\negthinspace)}}
 \put(-4,0){\rstrip{0}}         \put(204,0){\lstrip{0}}
 \put(-4,45){\rstrip{\ln{\frac54}}}     \put(204,45){\lstrip{.322}}
 \put(-4,139){\rstrip{\ln 2}}   \put(204,139){\lstrip{1}}
 
 \put(65,130){\rstrip{\scriptstyle{p=0}}}
 \put(100,80){\rstrip{\scriptstyle{p=1/4}}}
 \put(110,50){\rstrip{\scriptstyle{p=1/2}}}
 \put(115,25){\rstrip{\scriptstyle{p=3/4}}}
 \put(118,4){\rstrip{\scriptstyle{p=1}}}
 
 \put(100,110)
        {\makebox(0,0){$\displaystyle \mutu=H(\pbar\,\qbar)-\qbar H(p)$}}
 \end{picture}}
{Mutual information $\mutu$ of the ``used tape'' channel, plotted
as a function of the punch density $q$ for various values of
the current hole density $p$ treated as a parameter. The maximum
of each curve is marked.}

The \tdf{capacity} $C$ of the channel is the maximum of $\mutu$ over all
possible values of $q$ (or, equivalently, of $\pp$).
	By equating to zero the derivative of $\mutu$ with respect to
$\pp$,
 $$
	\der\pp{\mutu} = \LN\frac\pp\ppbar+\frac{H(p)}\pbar=0,
% Der qM=\pbar\LN\frac{\pbar\,\qbar}{1-\pbar\,\qbar}+H(p),
 $$
	one finds that this maximum occurs at
 \Eq[mono_qhat]{
	\qhat = 1-\frac1{\pbar(e^{H(p)/\pbar}+1)},\ \hbox{or}\
	\pphat=\frac1{e^{-H(p)/\pbar}+1},
}
	where $\mutu$ attains the value
 \Eq[mono_capacity]{
	C = \ln(e^{-H(p)/\pbar}+1) = \LN \pphat,
}
	as plotted in \fig{mono_capacity}. In particular, for $p=1/2$,
 $$
	\qhat=\frac35,\ \pphat=\frac45,\ \hbox{and}\
	C=\ln\frac54\approx.322\ \bit.
 $$

\Fig[mono_capacity]{%
 \begin{picture}(248,156)(-24,-14)
 \small
 \put(0,0){\includegraphics{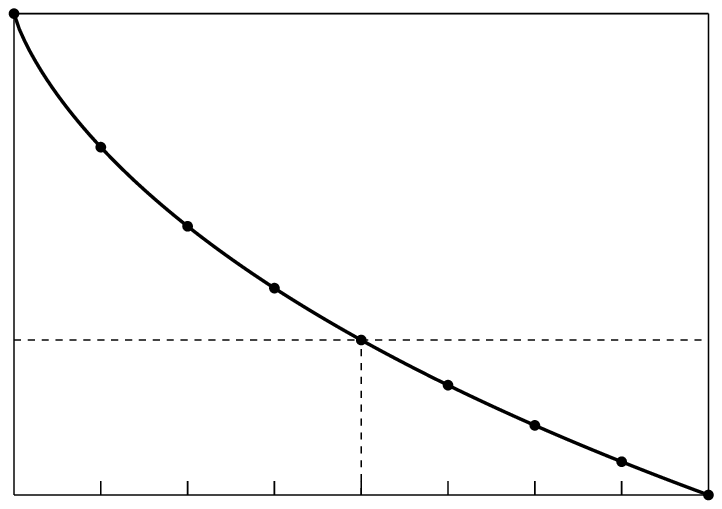}}
 \put(224,-10){\rstrip{\hbox{\normalsize$p\to$}}}
 \put(0,-8){\cstrip{0}}
 \put(50,-8){\cstrip{\frac14}}
 \put(100,-8){\cstrip{\frac12}}
 \put(150,-8){\cstrip{\frac34}}
 \put(200,-8){\cstrip{1}}
 
 \put(-24,120){\lstrip{\hbox{\normalsize$\above0\uparrow C$}}}
 \put(201,129){\lstrip{(\negthinspace\times\hbox{bit}\negthinspace)}}
 \put(-4,0){\rstrip{0}}         \put(204,0){\lstrip{0}}
 \put(-4,45){\rstrip{\ln{\frac54}}}     \put(204,45){\lstrip{.322}}
 \put(-4,139){\rstrip{\ln 2}}   \put(204,139){\lstrip{1}}
 
 \put(120,110){\makebox(0,0){$\displaystyle{C=\ln(e^{-H(p)/\pbar}-1)}$}}
  \end{picture}}
 {Channel capacity of used tape, $C$, as a function of the current hole
density $p$; the dots match those of \protect\fig{mono_mutu}.}

\Fig[mono_qhat]{%
 \begin{picture}(248,216)(-24,-12)
 \small
 \put(0,0){\includegraphics{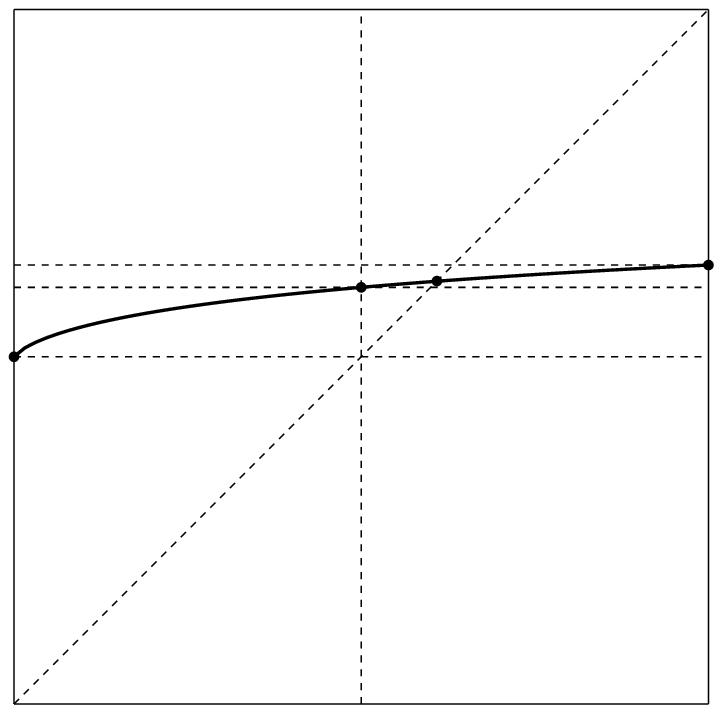}}
 \put(224,-10){\rstrip{\hbox{\normalsize$p\to$}}}
 \put(0,-8){\cstrip{0}}
 \put(100,-8){\cstrip{1/2}}
% \put(121.8,-8){\cstrip{\kappa}}
 \put(200,-8){\cstrip{1}}
 
 \put(-16,180){\lstrip{\hbox{\normalsize$\above0\uparrow\qhat$}}}
 \put(-4,0){\rstrip{0}}         \put(204,0){\lstrip{0}}
 \put(-4,100){\rstrip{1/2}}     \put(204,100){\lstrip{.5}}
 \put(-4,115){\rstrip{3/5}}     \put(204,115){\lstrip{.6}}
% \put(-4,122.5){\rstrip{\kappa}}\put(204,122.5){\lstrip{.609}}
 \put(-2,130){\rstrip{1\hbox{$-$}1/e}}  \put(204,130){\lstrip{.632}}
 \put(-4,200){\rstrip{1}}       \put(204,200){\lstrip{1}}
 
 \put(50,150){\makebox(0,0){$\displaystyle\qhat=1-\frac1{\pbar(e^{H(p)/\pbar}+1)}
$}}
 \end{picture}}
 {Plot of the optimal punch density $\qhat$ vs current hole density $p$. The
distinguished points are discussed in the text. That lying on the $\qhat=p$
line is discussed in \protect\sect{tape wars}.}

\Sect{Cooperation and competition}

For sake of contrast with the current context of selfish, independent
utilization of the tape by each successive party, in the following two
subsections we shall briefly discuss the possibilities of cooperation and
competition.

\subsection{You shall receive an hundredfold}

In two successive selfish transmission stages starting from virgin tape,
Alice got 1 bit's worth of message out of each cell and Bob .322 bit, for a
total of 1.322 bit. By collaborating, they could do much
better\cite{rivest-worm,maier}. In fact, if Alice and Bob worked in concert,
with a very simple code they could each get two bits' worth of message out of
every three cells, for a total of $4/3\approx1.333$ bit/cell; with long block
codes, they could get up to about 1.55 bit/cell. The advantages of
collaboration show up even better when one can plan ahead a long series of
transmission stages with a long length of tape: in this situation, the
cumulative amount of message worth one can get out of an $n$-cell length of
tape grows as $n\ln n$; therefore, the amount per unit length is unbounded!

\subsection{Tape wars averted}\sectlabel{tape wars}

We have seen that, if the original tape was punched at a density $p$ by
Alice, with an attendant rate $H(p)$ for her message, then Bob
can achieve his channel capacity $C(p)$ as in \eq{mono_capacity} by punching
at density $\qhat(p)$ as in \eq{mono_qhat}.  In the process, Alice's original
message is, of course, degraded. In fact, if Alice tries to read back her
message, she will find it contaminated by the same amount of one-way noise as
if it had gone through the channel described by exchanging $q$ and $p$ in
\fig{mono_channel-diagram} and table \eq{mono_channel-matrix}.

Suppose now that Alice, realizing that her tapes are going to be reused (or
\emph{concurrently} used---since, as we have seen, the two punching
operations commute) by Bob, decides to encode her next batch of tapes so as
to make her messages readable even after an anticipated punching by Bob at
density $q$. According to \eq{mono_qhat} and \fig{mono_qhat}, as a preventive
measure she will have to shift her punch density $p$ to a higher value than
1/2, thus achieving a lower rate but greater resistance to Bob's
tampering. When Bob realizes that, he will be forced to shift \emph{his} punch
density $q$ to a higher value---and so forth.

This is not a zero-sum game: as the arms race unfolds, each party will end up
storing progressively less information on the tape. Will the race lead to the
mutual destruction of information capacity?  Fortunately, the curve of
\fig{mono_qhat} intersects the line $\qhat=p$ and there has a slope less than
1. Thus, the race converges to a stable point (with $q=p\approx 0.609$),
where each party achieves an effective storage capacity of $\approx.240$
bit/cell.

The sum of the two capacities---and these are \emph{coexisting} capacities,
with both messages readable at the same time!---is about 0.48 bit/cell, to be
compared with the 1 bit/cell Bob and Alice could have achieved by
``space-sharing'' the tape (e.g., one cell for Alice, one for Bob, and so
forth).  Thus, the attempt by the two parties to concurrently use the
monotonic-write tape, performed in a selfish but rational way, results in an
overall loss of storage capacity that is substantial but not crippling.

It must be noted that, even though at equilibrium they are in a symmetric
situation, Alice and Bob cannot use the \emph{very same} block code to encode
their messages on the tape. To avoid interference the two codes must be
practically uncorrelated or ``mutually orthogonal''; this is always possible
with long enough block codes.

\Sect[dialogues]{Intercom dialogues}

We introduce the issue of \emph{tape equivalence} by means of two dialogues.
The \tdf{length} $\ell$ of a piece of tape is the number of cells it
contains.  Because of the canonical distribution of both holes and punches
(\sect{orientation}), the overall capacity of a tape of length $\ell$ is
$\ell$ times the capacity of a single cell, and similarly for the mutual
information.

%\def\actor#1#2{%		\actor<tag><full name>
%  \newcommand{\cs #1\endcs}[1][]{\par\smallskip\noindent\tsc{#2}##1.\enskip}}
%\actor{bob}{Bob}
%\actor{sue}{Sue}

\def\actor#1#2{\par\smallskip\hangindent1em\noindent\tsc{#1}\if*#2*\else,\ \emph{#2}\fi:\enskip}

\newcommand{\bob}[1][]{\actor{Bob}{#1}}
\newcommand{\sue}[1][]{\actor{Sue}{#1}}
\newcommand{\wil}[1][]{\actor{Willie}{#1}}

\subsection*{Dialogue 1}

{\em Bob is now an old and stingy facilities officer at Caltech. He can no
longer see the individual holes on the tape---his vision is blurred---and he
wouldn't any longer know how to start writing a block code.  All he
cares about is tape as a bulk commodity, and getting the most out of it. He
is assisted by Sue, who physically handles the tape and knows how to devise
appropriate block codes. Sue has standing instructions to recycle paper tape
to the best of her capabilities and not to bother Bob with details.}

 \smallskip
 \bob[on the intercom] Sue, we have to send a million-bit message to
MIT. Get a piece of tape.
 \sue[from the mail room] I've got here a reel of tape with an overall
capacity of one million bits. [\emph{She doesn't tell Bob whether that's a
thousand feet of virgin tape, or perhaps ten thousand feet of heavily used
tape.}]
 \bob Good! Here is the message. Don't waste any capacity, and make sure you
get the tape back from MIT so we can reuse it! By the way, what will be the
capacity left on the tape after this message? I want to enter it as an asset in
my inventory sheet.
 \sue That will be 333,000 bits.
 \bob So, using this tape at capacity will leave it ``shrunk'' to .333
of its previous capacity. Well, one third left is better than nothing!
 % Note: A shrink factor of 1/3 means that p=.1096. qhat=.5611, p'=.59585
 % The next shrink factor will be 1/2.8076
 \bob[a week later] Sue, here is another message for MIT. Since it happens to
be 333,000 bits long, let's use the tape you got back from them. [\emph{We
assume that, after decoding a message, MIT does not keep a record of the
detailed hole pattern received. That might be used for improving transmission
efficiency, but at substantial storage cost.}] What will be the capacity left
on the tape after this message?
 \sue That will be about 119,000 bits.
 \bob[punching keys on a calculator] Hey, this time, it will only shrink to
119,000/333,333=.357 of its pre-transmission capacity [\emph{with an
accusing tone}] Are you sure you made the best use of my tape last time?
 \sue[chuckling] Cool off, Bob! Every housewife knows that used tape shrinks
less!  In fact, really ripe tape only shrinks to $1/e\approx .368$ of its
previous capacity upon each usage.
 \bob And brand new tape?
 \sue New tape is the worst! It will shrink to $\ln5/\ln2-2\approx.322$ of its
previous capacity. Here's the whole picture! (\fig{shrink-factor})

\Fig[shrink-factor]{%
 \begin{picture}(248,92)(-24,-12)
% 1/e=0.3678
% ln5/ln2 - 2 = .322
 \small
 \put(0,0){\includegraphics{fig/shrink-factor.eps}}
 \put(224,-10){\rstrip{\hbox{\normalsize$p\to$}}}
 \put(0,-8){\cstrip{0}}
 \put(64,-8){\cstrip{.9}}
 \put(107,-8){\cstrip{.99}}
 \put(137,-8){\cstrip{.999}}
 \put(160,-8){\cstrip{.9999}}
 \put(200,-8){\cstrip{1}}
 
 \put(-12,32){\rstrip{\hbox{\normalsize$\displaystyle\above0\uparrow s$}}}
 \put(-4,0){\rstrip{0}}                  \put(204,0){\lstrip{0}}
 \put(-4,63){\rstrip{\scriptstyle\frac{\ln5}{\ln2}-2}}
                                         \put(204,64.5){\lstrip{.322}}   
 \put(175,69.5){\cstrip{\scriptstyle\cdots}}  \put(198,69.5){\rstrip{\scriptstyle
\infty}}
 \put(-4,74.5){\rstrip{1/e}} \put(204,73.5){\lstrip{.368}}       
 \end{picture}}
 {``Shrinkage coefficient'' $s=C(\pphat)/C(p)$ for successive full-capacity
usages (labeled 1, 2, \ldots, $\infty$) of an initially virgin tape. As the
tape gets more thoroughly used, the shrinkage coefficient rapidly converges
to $1/e$.}

\medskip

The two physical parameters of a piece of tape, namely, its length $\ell$ (in
cells) and its current hole density $p$, completely characterize its
``response''---in terms of amount of information transmitted and capacity
left---upon each successive usage, including usages with a punch density
$q<\qhat$ (where some of the capacity is saved for later) or $q>\qhat$ (where
some capacity is wasted), according to equations \eq{pprime}, \eq{mono_mutu},
and \eq{mono_capacity}.

In particular, the capacity of a piece of tape of length $\ell$ is $\ell
C(p)$ (cf.~\eq{mono_capacity}). This can be thought of---if we measure
capacity in bits (see \sect{notation})---as the \tdf{reduced length} of the
tape---i.e., the number of cells of virgin tape having the same overall
capacity. Bob would have been delighted to find that two pieces of tape
having the same reduced length are completely equivalent for
information-transmission purposes. Such an equivalence principle would allow
him to characterize a piece of tape by means of a single
information-theoretical parameter---the reduced length---rather than the two
physical parameters $\ell$ and $p$, and greatly simplify his inventory
bookkeeping.

If such an equivalence held, then, as a specific consequence, the shrinkage
coefficient of Dialogue 1, defined as
 $$
	s(p)=\frac{C(\pphat)}{C(p)},
 $$
	would be independent of $p$. Unfortunately, as we have seen in the
dialogue, this is only approximately true (\fig{shrink-factor}). We'll return
to this problem, with better tools, in \sect{strong-principle}.

\subsection*{Dialogue 2}

{\em Sue is on vacation. Her temporary replacement, Willie, is being
indoctrinated by Bob about the need to conserve tape.  To test his coding
capabilities, Bob chooses a spool of tape just like the one he gave Sue the
first time.}

\smallskip

\bob Here is a length of used tape, Willie, and a million-bit message to be
sent to MIT. Please transmit the message as efficiently as you can.

\wil Is it urgent?

\bob Not, really. Take your time, but do a good job!

\bob[a month later] Well, did you get the tape back from MIT?

\wil Here it is!

\bob What's its capacity now?

\wil 580,000 bits, more or less.

\bob What? It only shrank to .580 of its original length? How did you manage
that?

\wil You know, haste makes waste. So I first encoded only a small fraction of
the message on the tape, using a very low punch density. MIT decoded that,
wrote it down, and sent back the tape. Then I encoded on the same tape
another increment of the message, sent it to MIT, and so on. The tape
must have gone back and forth twenty times!

\bob In the limit of an infinite number of infinitesimal
increments, how much information could you transmit in this way?

\wil Starting from virgin tape, about 2.37 bits/cell
(precisely, $\frac{\pi^2}{\ln2}$).

\bob That's amazing!

\wil And, of course, at any intermediate moment the transmission ``mileage''
already used plus that which is still left on the tape equals a
constant---provided you always travel very slowly.

\bob I got it! Your ``mileage left'' is the \tdf{effective length} I was
looking for. No matter how different they look physically, two pieces of tape
(say, one short and fresh and the other long and stale) having the same
effective length are equivalent for information transmission purposes.

\wil Slow down, Bob! That is true only as long as you use them up
\emph{slowly}. By comparing Sue's performance with mine, you realize that,
when one tries to cram onto a tape a substantial fraction of its channel
capacity at once, there are losses by ``friction'', as it were.\footnote
	{\label{impedance-mathc}This behavior is qualitatively similar to
that of mechanical systems. Consider a battery of internal resistance $R$
connected to a load of impedance $r$. It will be convenient to use the
normalized variable $p=1/(1+r/R)$, which goes from 0 to 1 as $r/R$ goes from
$\infty$ to 0.  The maximum \emph{power transfer} occurs when $p=1/2$ (i.e.,
$r=R$); in this case, half of the energy is dissipated by friction in $R$. As
$p\to0$, energy is transfered to the load more slowly but less of it is
wasted by friction. (As $p\to1$, one gets less power out \emph{and} wastes a
greater fraction of the energy.)  Indeed, to an untrained eye the power
transfer curve $2p(1-p)$---an inverted parabola---is hard to tell apart from
the binary entropy curve $H(p)$.}
	Well, one can tell the difference between fresh tape and well-worn
tape by the fact that the former exhibits \emph{just a little more friction}
than the latter.

\Sect{Weak equivalence}

Let us explore in more detail what Bob discovered with Willie's help.

Suppose that we start with virgin tape and record on it a small amount $d\info$
of information by punching it at a very low density.  We ship the tape
but ask the recipient to send it back to us after reading the
message. We then record on this ``slightly used'' tape an additional small
amount of information,\footnote
	{For this, we need, of course, a very long block code.}
further increasing the hole density. We continue in
this way, sending one after the other a large number of messages each having
a small information contents, until the tape is completely filled with holes.
If at each stage the encoding is done optimally, what is the cumulative
information $\int d\info$ of the messages we sent?

Assume that at a generic stage of this process we start with a hole density
$p$ and increase it to $\pp=p+dp$ by issuing punch commands with a
probability $dq$ per cell.  The channel diagram is the same
as \fig{mono_channel-diagram}, but with input and output probabilities
as in \fig{incr_channel-diagram}.

\Fig[incr_channel-diagram]{%
 \vbox{ % the \vbox is to keep the freedom of vmode until business starts
 % in case we are called from within display mode
 % \prop is vertical struts, fixed depth greater than any expected
 \def\prop#1{\vrule width 0pt height #1pt depth\Foot pt}
 \def\Foot{10}
 \def\afterheading{8}
 \def\adjafterheading{5}        %afterheading minus 3

 \def\ChannelLength{72}         % nominal length of arrows
 \def\ChannelHeight{48}         % height from top dot to bottom dot
 \def\AdjChanHeight{38}         % ChannelHeight minus Foot
 \def\vl{33}                    % 1/2 ChannelLength - 3

 \hbox{%
 \begin{array}[t]{cc@{\ }c@{}}
        P(x)    & x     &               \prop{0}\\
        \overbar{dq}& \hbox{spare}      &\bullet        \prop{\afterheading}\\
        dq      & \hbox{punch}  &\bullet        \prop{\AdjChanHeight}
 \end{array}%
 \vtop{\hbox{\prop{0}}
   \hbox{\vrule width 0pt height\adjafterheading pt
      \begin{picture}(72,0)(-36,0)
        \def\CP{-12}
        \def\AA{0}      \def\AAtag{6}   % approx 6pt higher than line
        \def\AB{-24}    \def\ABtag{-10} % approx 6pt higher than line
        \def\BB{-48}    \def\BBtag{-42} % approx 6pt higher than line
        \put(\CP,\AAtag){\makebox(0,0){$\pbar$}}
        \put(\CP,\ABtag){\makebox(0,0){$p$}}
        \put(\CP,\BBtag){\makebox(0,0){$1$}}
        \put(0,\AA){\vector(1,0){\vl}}\put(0,\AA){\line(-1,0){\vl}}
        \put(0,\AB){\vector(3,-2){\vl}}\put(0,\AB){\line(-3,2){\vl}}
        \put(0,\BB){\vector(1,0){\vl}}\put(0,\BB){\line(-1,0){\vl}}
      \end{picture}}}%
 \begin{array}[t]{@{}c@{\ }cc}
                & y     & P(y)          \prop{0}\\
        \bullet & \hbox{blank}  & \pbar\overbar{dq}\prop{\afterheading}\\
        \bullet & \hbox{hole}   & p\overbar{dq}+1dq\prop{\AdjChanHeight}
 \end{array}%
 }}}
 {Channel diagram for used tape, with input and output probabilities
corresponding to incremental use of the channel capacity.}

The hole density increment is $dp=\pbar dq$, as the blanks, which appear with
density $\pbar$, are turned into holes with probability $dq$, while the
holes, with density $p$, remain unaffected.  The mutual information of this
infinitesimal punching operation, calculated from \eq{mono_mutu} using $dq$
in place of $q$, is
 \begin{align}
	d\info &=H(p+dp)-\frac{\overbar{p+dp}}\pbar H(p)\\
	   &= \left(\LN\frac p\pbar+\frac{H(p)}\pbar\right)dp
		= \frac{\LN p}\pbar dp.
 \end{align}
	The indefinite integral of the integrand in the last expression is
 $$
	\int\frac{\LN p}{1-p}dp = -\mathrm{Li}_2(1-p),
 $$
	where $\mathrm{Li}_2$ is the \emph{dilogarithm} function.\footnote
	{This is one of the \emph{polylogarithm} functions, defined by
$\mathrm{Li}_n(z)=\sum_{k=1}^\infty\frac{z^k}{k^n}$.}
	Thus, the \tdf{effective capacity} of a tape of hole density $p$,
i.e., the total amount of information that can be transmitted via it in
successive small increments until all holes have been punched up, is
 \Eq[mono_effective]{
	Q(p)= \int_p^1d\info=-\Bigl.\mathrm{Li}_2(1-x)\Bigr|_p^1=\mathrm{Li}_2(\pbar),
}
	as plotted in \fig{mono_effective} (compare with the qualitatively
similar behavior of $C$, in \fig{mono_capacity}); for virgin tape ($p=0$),
the effective capacity is $\mathrm{Li}_2(1)={\pi^2/6}$ (cf.\ \cite{wolf}).
Note that, by \eq{mono_effective},
 $$
	d\info=-dQ.
 $$
	Since $Q$ is a \emph{function of state} of the tape (i.e., it depends
only on its state and not on the specific sequence of operation that led to
that state), $d\info$ is an \emph{exact} differential.

\Fig[mono_effective]{%
 \begin{picture}(248,156)(-24,-14)
 \small
 \put(0,0){\includegraphics{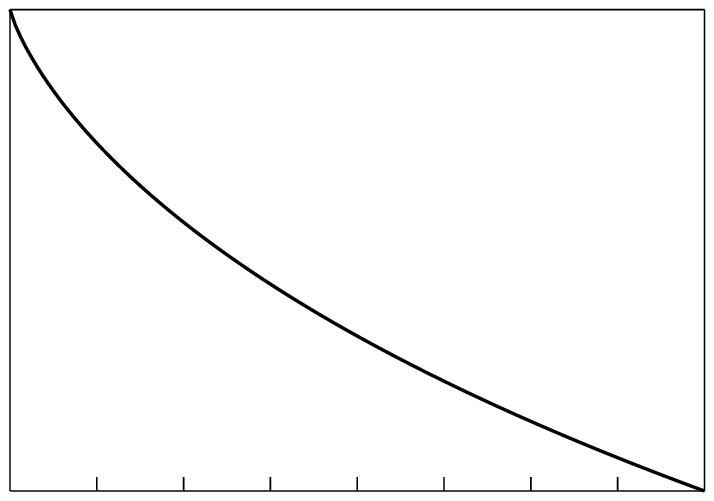}}
 \put(224,-10){\rstrip{\hbox{\normalsize$p\to$}}}
 \put(0,-8){\cstrip{0}}
 \put(50,-8){\cstrip{\frac14}}
 \put(100,-8){\cstrip{\frac12}}
 \put(150,-8){\cstrip{\frac34}}
 \put(200,-8){\cstrip{1}}
 
 \put(-24,92){\lstrip{\hbox{\normalsize$\above0\uparrow Q$}}}
 \put(201,129){\lstrip{(\negthinspace\times\hbox{bit}\negthinspace)}}
 \put(-4,0){\rstrip{0}}         \put(204,0){\lstrip{0}}
 \put(-1,138){\rstrip{\pi^2\negthinspace/6}}    \put(204,139){\lstrip{2.37}}

 \put(120,110){\makebox(0,0){$\displaystyle
                {Q=\mathrm{Li}_2(\pbar)}$}}
 \end{picture}}
 {Effective capacity $Q$ as a function of the current hole density $p$.}

The above quantities are on a per-cell basis. Let us define the
\tdf{effective length} (cf.\ Dialog 2) of a piece of tape of length $\ell$
and hole density $p$ as $\lambda=\ell Q(p)$. If by a sequence of small
incremental messages we transmit an amount of information $I$ per cell, and
thus a total amount $S=I\ell$ for the entire piece of tape, the new effective
length will be $\lambda'=\ell(Q-I)$. The corresponding shrinkage
coefficient\footnote
	{This quantity is analogous to but distinct from the shrinkage
coefficient of Dialogue 1, which is a ratio of channel capacities.}
	will be
 $$
	\frac{\lambda'}{\lambda}=1-\frac IQ=1-\frac S\lambda,
 $$
	which is \emph{independent} of the physical parameters $\ell$ and $p$
and depends only on the ratio between two information-theoretical quantities,
i.e., the total amount $S$ of information transmitted and the effective
length $\lambda$ of the tape.  We shall call this the \tdf{weak equivalence
principle} for monotonic-write media.

 \Figwide[mufam]{%
 \begin{picture}(448,188)(-24,-12)
% (6/pi^2)ln2 = .42138
% 1/e=0.3678

 \small
 \put(0,0){\includegraphics{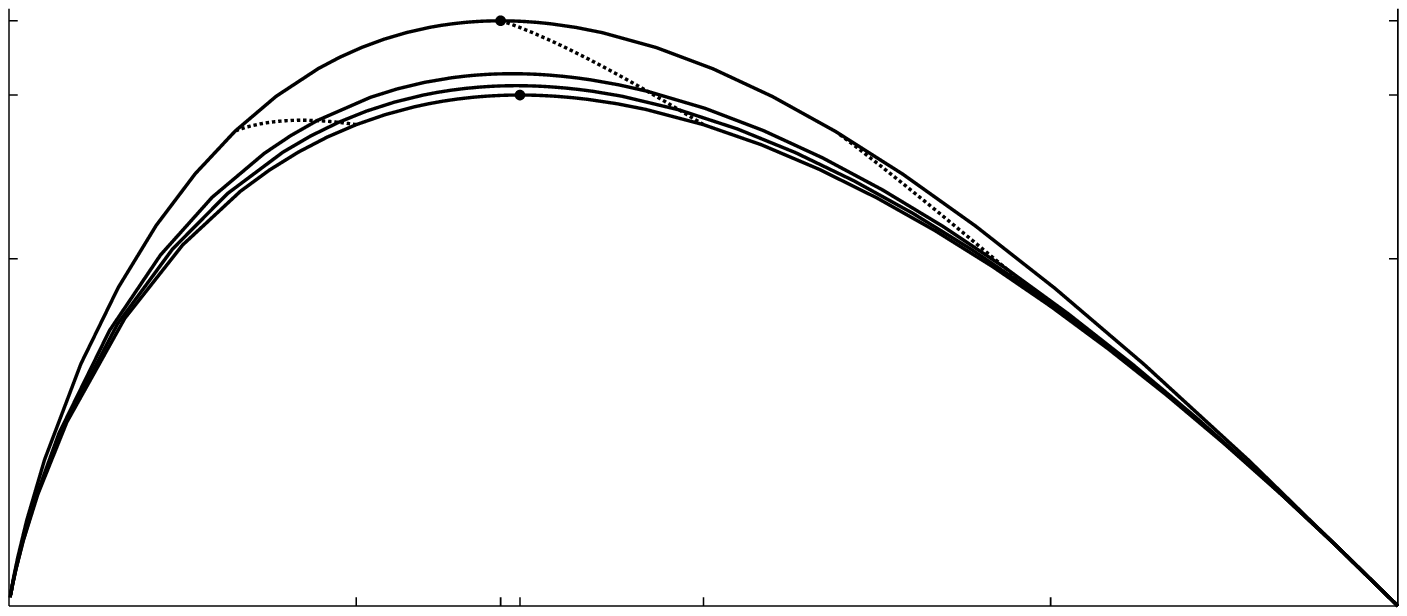}}
 \put(424,-10){\rstrip{\hbox{\normalsize$\sigma\to$}}}
 \put(0,-8){\cstrip{0}}
 \put(100,-8){\cstrip{\frac14}}
% \put(141.6,-8){\cstrip{\kappa}}
 \put(140,-8){\cstrip{\kappa}}
% \put(147.2,-8){\cstrip{\frac1 e}}
 \put(148.5,-8){\cstrip{\frac1 e}}
 \put(200,-8){\cstrip{\frac12}}
 \put(300,-8){\cstrip{\frac34}}
 \put(400,-8){\cstrip{1}}
 
 \put(-12,64){\rstrip{\hbox{\normalsize$\above0\uparrow\mu$}}}
 \put(-4,0){\rstrip{0}}                  \put(404,0){\lstrip{0}}
 \put(-4,100){\rstrip{1/4}}              \put(404,100){\lstrip{.250}}
 \put(-4,146){\rstrip{1/e}} \put(404,147){\lstrip{.368}}         
 \put(-1,168){\rstrip{\frac6{\pi^2}\ln2}} \put(404,168){\lstrip{.421}}   
 
 \put(4,6){\lstrip{q=1}}
 \put(76,99){\cstrip{q=\frac34}}\put(76,103){\vector(0,1){35}}
 \put(160,123){\cstrip{q=\frac12}}\put(160,127){\vector(0,1){35}}
 \put(250,90){\cstrip{q=\frac14}}\put(250,94){\vector(0,1){35}}
 \put(390,6){\rstrip{q=0}}
 
 \put(90,164){\cstrip{p=0}}
 \put(112,155){\cstrip{\scriptstyle1/2}}
 \put(132,159){\lstrip{\scriptstyle3/4}}\put(130,159){\vector(0,-1){9}}
 \put(125,140){\cstrip{p=1}}
  \end{picture}}
 {Mutual information per unit of effective length, $\mu$, as a function of
the shrinkage coefficient $\sigma$, for different values of the initial hole
density $p$. The dotted lines represent loci of equal values for the running
parameter $q$. For any value of $p$, the maximum of $\mu$ represents the
corresponding channel capacity given in units of effective length. The
maximum of $\mu_0$ occurs at $\kappa=\frac12(1-\frac6{\pi^2}\ln^2 2)$. For
utilization rate below capacity, the same mutual information can be obtained
with two different values of shrinkage, one corresponding to rational usage
of the tape and the other to needlessly wasteful usage.}

\Sect[strong-principle]{Strong equivalence}

Let us now explore in more detail what Bob discovered with Sue's help.

Whether we intend to utilize a piece of tape incrementally, as in Dialogue 2,
or in discrete installements, as in Dialogue 1, the \emph{effective length}
$\lambda$ defined above provides a more natural measure of a tape's information
capacity than the \emph{reduced length} introduced in Dialogue 1.

Armed with this measure, let us now turn our attention from the special case
of the limit of an infinite sequence of infinitesimal messages to the general
case of finite-size messages, where the weak principle is not applicable.

Our goal is to eliminate the physical parameters $p$ and $q$ between
equations \eq{mono_mutu} and \eq{mono_effective}, and thus write a relation
directly between (a) the effective length $\lambda$ of a piece of tape before
the transmission of a message, (b) the effective length $\lambda'$ after the
transmission, and (c) the amount $S$ of information conveyed by the
message. If such a relation exists, it may be assumed to be of the form
 $$
	f(\lambda,\lambda',S)=0
 $$
	and, since we are assuming a canonical hole distribution before and
after punching, it must satisfy the scaling property
 $$
	f(a\lambda,a\lambda',aS)=0\quad\hbox{for any}\ a.
 $$
	Setting, as a special case, $a=1/\lambda'$, we obtain a relation
between two variables
 $$
	g(\sigma,\mu)=f(\sigma,1,\mu)=0,
 $$
	where
 $$
	\sigma=\frac{\lambda'}\lambda=\frac{Q(\pp)}{Q(p)}\quad \hbox{and}\quad
	\mu=\frac S\lambda=\frac{\info(p,q)}{Q(p)}.
 $$
	The variable $\mu$---which is the mutual information for a given
stage of utilization of the tape---can be thought of as the information rate
per unit of effective length of the tape, and $\sigma$ as the shrinkage
coefficient attendant to that stage.

Since the variables $\sigma$ and $\mu$ depend on two parameters, $p$ and $q$,
we cannot \emph{a priori} expect to eliminate \emph{both} parameters when
solving for $\mu$ with respect to $\sigma$. However, for a given initial hole
density $p$ treated as a fixed parameter, we can eliminate just $q$ and write
 $$
	\mu=\mu_p(\sigma).
 $$
	The result of this elimination, performed numerically for different
values of $p$, are shown in \fig{mufam}, which also shows the values of the
eliminated parameter $q$ on the $\mu(\sigma)$ curves.

Paralleling the weak equivalence principle of the previous section---which
states that tapes having the same effective capacity are indistinguishable at
slow utilization rates---a \tdf{strong equivalence principle} would be one
that is valid for any rate of utilization of the tape at any transmission
stage, from an infinitesimal hole-density increment ($q$ close to 0) to gross
overpunching ($q$ close to 1). I don't know whether it is more surprising
that, strictly speaking, punched tape does \emph{not} obey a strong
equivalence principle, or that, after all, it turns out to do so \emph{to a
very good approximation}.  In fact, as is clear from \fig{mufam}, after
eliminating $q$ between $\mu$ and $\sigma$ some dependence on $p$ remains,
but this dependence is slight in any case and rapidly vanishes as $p$
approaches 1.  Intuitively, the one-parameter family of curves of
\fig{mono_mutu} nearly collapses---when expressed in terms of a more natural
set of variables---onto a \emph{single curve} (\fig{mufam}).

The curves $\mu_p(\sigma)$ all have slope $-1$ at $\sigma=1$; this is an
expression of the weak equivalence principle (i.e., for small $q$, the
effective length decreases by an amount equal to the amount of information
transmitted). They all have slope $\infty$ at $\sigma=0$, signifying that the
waste of effective capacity increases precipitously when one punches at a
density much greater than that needed for transmitting at channel capacity.

The worst-case departure of the $\mu_p$ curves from the limiting curve
$\mu_1=\lim_{p\to1}\mu_p$ occurs near the maximum bulge of the curves, and is
substantially the same as the departure of $s$ from its $1/e$ limit as
plotted in \fig{shrink-factor}. The curves $\mu_p$ are not likely to be
expressible in closed form; however, as is easy to prove, the limiting curve
$\mu_1$ is nothing but the familiar \emph{self-information} function
$\mu=\sigma\LN\sigma$.  To the same approximation as the strong equivalence
principle holds, \emph{this function gives the information-transfer
characteristics of punched tape (i.e., for any message, the capacity used by
it, that wasted, and that left after the message) over the tape's entire
utilization range}.

Let us remark that the self-information function appears in the limit also in
\fig{mono_mutu}. In fact, one can show that
 $$
	\lim_{p\to1}\frac{\info(p,q)}{C(p)}=e\qbar\LN\qbar.
 $$ 

\Sect{Conclusions}

A piece of randomly punched tape is described by two physical
parameters---its length $\ell$ and its hole density $p$.
	We have raised the question of whether the tape's behavior as an
information transmission commodity can be usefully characterized by a single
information-theoretical parameter---its \emph{effective length} $\lambda$.
We have concluded that this is the case
 \begin{list}{$\bullet$}{\compress}
	\item in the ``quasi-static'' limit of
slow utilization rate (\emph{weak equivalence principle});
	\item for any utilization rate (\emph{strong equivalence principle})
	\begin{list}{$-$}{\compress}
		\item \emph{exactly}, but only in the limit of already
heavily used tape, and
		\item \emph{approximately}---but with good accuracy even in
the worst case---over the whole range of previous and future uses of the
tape.
	\end{list}
 \end{list}

\Sect{Acknowledgements}

This research was funded in part by NSF (9305227-DMS) and in part by ARPA
through the Ultra Program (ONR N00014-93-1-0660) and the CAM-8 project (ONR
N00014-94-1-0662). I am indebted to Peter Elias, Matteo Frigo, and Mark Smith
for useful discussions, and to Ronald Rivest for some references.

\bibliographystyle{IEEE}

\begin{biography}{Tommaso Toffoli} Tommaso Toffoli received a Doctorate in
Physics from the University of Rome, Italy, in 1967, and a Ph.D.  in Computer
and Communication Science from the University of Michigan, Ann Arbor, in
1976.  In 1977 he joined the MIT Laboratory for Computer Science, eventually
becoming the leader of the Information Mechanics group. In 1995 he joined the
faculty of the Boston University ECE Department.\newline
	His main area of interest, namely Information Mechanics, deals with
fundamental connections between physical and computational processes.  He has
developed and pioneered the use of cellular automata machines, as a way of
efficiently studying a variety of synthetic dynamical systems that reflect
basic constraints of physical law, such as locality, uniformity, and
invertibility. Related areas of interest are: quantum computation;
correspondence principles between microscopic laws and macroscopic behavior;
quantitative measures of ``computation capacity'' of a system---as contrasted
to ``information capacity,'' and connections between Lagrangian action and
amount of computation.\newline
	A new initiative, Personal Knowledge Structuring, aims at developing
cultural tools that will help ordinary people turn the computer into a
natural extension of their personal faculties.

\end{biography}

\end{document}